   \documentclass[prl,aps,twocolumn,showpacs]{revtex4}
\usepackage[dvips]{graphicx}
\usepackage{dcolumn}
\newcommand{\be}{\begin{equation}}
\newcommand{\ee}{\end{equation}}
\newcommand{\bea}{\begin{eqnarray}}
\newcommand{\eea}{\end{eqnarray}}
\newcommand{\nn}{ \nonumber}

\begin{document}
\topmargin=-20mm

\title{Electrodynamics of Nearly-Ferroelectric Superconductors}

\author{Joseph L.Birman and Natalya A. Zimbovskaya}

\affiliation {Department of Physics, The City College of CUNY, New York, NY}

\date{\today}
 
\begin{abstract}
   We report here the frequency-dependent optical response of a nearly ferroelectric superconductor (NFE-SC), like Na$_{x}$WO$_3$, or n-SrTiO$_3$. From $\omega =0$, up to a critical frequency $ \omega_{c1}$, Meissner-like field extinction occurs as in a usual superconductor. In the "anomalous" region $\omega_{c1}<\omega<\omega_{TO}$ (the TO frequency of the NFE) the field propagates as in a dielectric. For $\omega >\omega_{TO}$ Meissner damping recurrs, then at higher $\omega$,
dielectriclike propagation is found. New optical anomalies (peaks in transmission) are predicted for a film of NFE-SC below the soft mode frequency $\omega_{TO}$. These effects are due to including the contribution of the NFE displacement current in the solution of the coupled Maxwell, lattice and London equations. Quantitative calculations are given for the NFE-SC n-SrTiO$_3 $ to illustrate this.
 \end{abstract}

\pacs{74.20.De; 74.25.Nf; 78.20.Bh}

\maketitle

\section{ I. Introduction}

This paper predicts novel effects in the electromagnetic response of a material which exhibits superconductivity (SC), and is in a "nearly-ferroelectric" (NFE) state. We use the description "nearly-ferroelectric" to mean a material which is a typical soft-mode phonon Cochran-Anderson system \cite{1}, with high-static dielectric constant. Materials in this class include  the sodium tungsten bronze Na$_x$WO$_3$ and n- (or p-) doped SrTiO$_3$ systems.

 Recently [2a] a sodium tungsten bronze Na$_x$WO$_3$ with $x \sim 0.05$ was reported to be a high temperature superconductor with $T_{c} \sim 90K$. It is noteworthy that for $0.1 < x < 1$, Mattheis and collaborators long ago reported superconductivity but with $T_{c} \sim 3-5K$ [2b]. Also Mattheis and Wood [2b] first reported ferroelectricity in the host WO$_3$ system. Another material, n- (or p-) doped SrTiO$_3$ exhibits
superconductivity, with $T_{c}\sim 1-3K$ \cite{3}. The host SrTiO$_3$ is known to be a "nearly-ferroelectric" material, where the static dielectric constant $\varepsilon (0)$ has been measured as $\sim 10^{4}$ in the temperature range where superconductivity occurs \cite{4}. In addition to
these two materials, Weger and collaborators have recently proposed that in the high temperature cuprate superconductors, a new multicritical point occurs due to an underlying nearly-ferroelectric instability which renormalizes the electron-phonon and electron-electron interaction and
enhances $T_{c}$ \cite{5,6}.

In the present work we will assume the ''host'' ionic lattice is a soft-mode NFE, with high dielectric
coefficient $\varepsilon (0)$. To clarify the scope of our work we also assume that all the carriers injected in the host by doping are fully condensed into $ s$-wave Cooper pairs in
that lattice, so there are no "free" electrons. This assumption is consistent with earlier work on such systems by Cohen and Koonce \cite{3} who presented a strong-coupling theory applied to superconductivity in SrTiO$_3$ (see also Ref. \cite{4}). Hence the free charge density $\rho_{e} =0$, and free current density ${\bf J}_{e} =0$.

 In our present phenomenological approach, we solve the coupled equations for: a) the transverse electromagnetic field (Maxwell equations); b) soft-mode lattice vibrations (Born-Huang equations); c) the superconducting electrons in London (local) approximation. We report dispersive structure, when the electromagnetic frequency is tuned through the resonant frequency region near the basic $ \omega_{T0}$ and $\omega_{L0} $ frequencies of the soft mode. The theory is applied to a
semiinfinite medium, and to a film in order to calculate the reflectivity and transmittivity (impedance). We illustrate the results quantitatively, using experimental data for superconducting doped SrTiO$ _3 ,$ which is one of the materials for which all the needed lattice, electronic and
superconducting data are available.

 We use London local electrodynamics for the superconducting sector(SC), rather than a nonlocal BCS or Pippard approach. These NFE-SC materials are all type II $ (\lambda_L > \xi $ where $ \lambda_L $ is the London penetration depth, and $ \xi $ is the coherence length) and in some cases they are strongly type II $ (\lambda_L >> \xi ) $ [7]. It is well known that London electrodynamics is valid in these cases. We also do not
consider in the present work effects due to vortex creation in these systems.

\section{II. Model And Dispersion}

At the outset we note that all the relevant dynamical fields considered, are transverse. These are the electromagnetic fields, the lattice vibrations, and the London supercurrent.  Maxwell's equations for the macroscopic fields are:
   \bea 
{\bf \nabla \cdot B} = 0, && \qquad
{\bf \nabla \cdot D} = 0, 
    \nn \\ \nn \\
{\bf \nabla \times E} + \frac{1}{c} \frac{\partial \bf B}{\partial t} = 0, && \qquad
{\bf \nabla \times H} - \frac{1}{c} \frac{\partial \bf D}{\partial t} = 0.
                               \eea
 with the constitutive equations of the medium taken as 
         \begin{equation}
{\bf B = H} + 4 \pi {\bf M} \quad \mbox{and} \quad
{\bf D = E} + 4 \pi {\bf P} = \varepsilon (\omega) \bf E.
                               \end{equation} 
 For the host ionic lattice we assume a ''diatomic'' basis, with $\bf w$ the relative displacement vector of the $ (\pm) $ ions ($ \bf w = u_+ - u_- )$. The equations of motion of the lattice, omitting damping, are \cite{8}:
         \begin{equation}
\frac{d^2 \bf w}{d t^2} = b_{11} {\bf w} + b_{12} {\bf E}, \qquad 
{\bf P} = b_{21} {\bf w} + b_{22} {\bf E},
                               \end{equation} 
 where $ b_{ij} $ are frequency-dependent coefficients and the
electric field $ \bf E $ and the polarization $ \bf P $ are the same as in Eqns.(1) and (2). Now we seek  a plane wave solution proportional to $ \exp (i {\bf k \cdot r} - i \omega t) $ for the Eqs. (3) in the long-wave approximation. We follow Huang, and for these transverse waves, in the ionic medium with a single resonance frequency $ \omega_{T0} $ we obtain
the usual expression for the dielectric function
         \begin{equation}
\varepsilon (\omega) = \frac{\varepsilon_\infty(\omega_{L0}^2 - \omega^2)}{(\omega_{T0}^2 - \omega^2)}
                               \end{equation} 
 where the expressions for the coefficients $ b_{i j} $ were given in Refs. \cite{8}. In media such as the perovskites, bronzes, or cuprates we have to take account of the other 
(nonsoft) L0 and T0 modes by renormalizing $ \varepsilon_\infty $ to some ''effective'' value $ \varepsilon'_\infty $ \cite{9}, where $ \varepsilon'_\infty = \varepsilon_\infty \Pi_{i}' \omega_{L i}^2 \big / \omega_{Ti}^2,$  (the primed product over $''i''$ includes all oscillations except the soft mode) and we
suppose the frequency $ \omega $ is far below all frequencies $ \omega_{L i }$ and $ \omega_{T i }$ except the soft mode. In what follows we use Eq. (4) with $ \varepsilon_\infty $ replaced by $ \varepsilon'_\infty $. The generalized Lyddane-Sachs-Teller (LST) relation is [10]:
         \begin{equation}
\frac{\varepsilon_0}{\varepsilon_{\infty}} = \frac{\Pi_{i}    
\omega_{Li}^2 }{ \omega_{T i}^2}.
                               \end{equation}
 Note that all  oscillators, including $ i =0 $ (the soft
mode) are included in Eq. (5) (no prime on the product). Finally we take the London equation \cite{11} in the
form:
         \begin{equation}
{\bf \nabla \times M} = \frac{1}{c}{\bf J}_s  = 
- \frac{1}{c^2 \Lambda} {\bf A } =
- \frac{n_s e^2}{m^* c^2} \bf A.     
                         \end{equation} 
 Here $\bf A $ is the vector potential for the electromagnetic field:
         \begin{equation}
{\bf \nabla \times A = B}; \qquad 
- \frac{1}{c} \frac{\partial \bf A}{\partial t} = \bf E.      
                         \end{equation} 
 The London gauge has been used, and in Eq. (6) $n_s $ is the superconducting electron density,and $ m^* $ is the electron effective mass. From Eqns. (1),(2),(4),(6) and (7) we obtain the equation for the electric field
         \begin{equation}
{\bf \nabla \times \nabla \times E} = 
- \frac{1}{c^2} \varepsilon (\omega)
\frac{\partial^2 \bf E}{\partial t^2} - \frac{4 \pi}{c^2 \Lambda} \bf E.
                               \end{equation} 
 The field equations describe the coupled
electromagnetic field, the soft mode NFE oscillator, and the London supercurrent.

  \begin{figure}[t]
\begin{center}
\includegraphics[width=8cm,height=7.5cm]{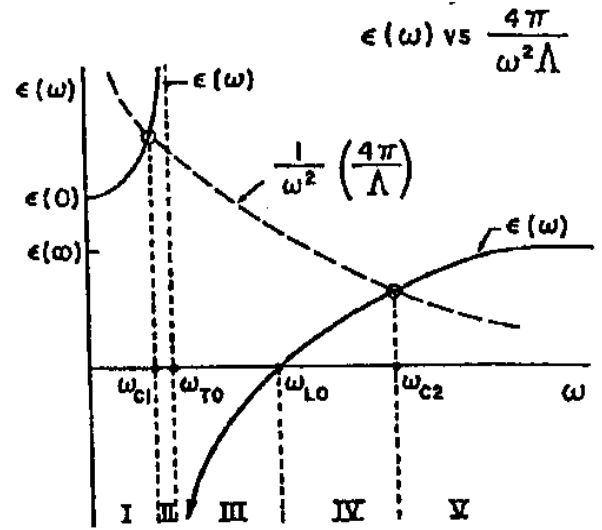}
\caption{ Solutions of the dispersion equation (9) for zero
damping. Frequency intervals $ \omega_{c1} < \omega < \omega_{T 0} $ and $ \omega > \omega_{c 2} $ correspond to real root of the Eq. (9).}
\label{rateI}
\end{center}
\end{figure}

We now look for a transverse (T) plane-wave solution for $ \bf E,$ and recalling that $ \bf \nabla \times \nabla \times E = - \nabla^2 E $ for such solutions, we obtain our new  result, the dispersion equation for $ T $ waves
         \begin{equation}
k^2 = \frac{\omega^2}{c^2} \varepsilon (\omega) - \frac{4\pi}{c^2 \Lambda}.      
                         \end{equation} 
 If the London term (last on the right-hand side) is absent, then Eq. (9) is the usual equation for a transverse phonon polariton \cite{8}; on the other hand, if the displacement term due to the polarization is absent (first term on the right-hand side) then the field attenuates in the distance  $\lambda_L^{-1}
= \sqrt{4 \pi/c^2\Lambda} $ where $ \lambda_L $ is the London penetration depth \cite{7}. Owing to the dispersive form (4) for $ \varepsilon (\omega) $, interesting physics emerges as $ \omega $ is increased from zero. Depending on the sign of the right-hand side of Eq. (9), $ k $ will either be real and electromagnetic waves will propagate, or imaginary so the wave will damp with frequency dependent penetration depth $ \lambda_L^* $ where
         \begin{equation}
\lambda_L^*  = \lambda_L  /\big (\big |1 -  \varepsilon (\omega) \lambda_L^2 {\omega^2}/{c^2} \big | \big
)^{1/2}.      
                        \end{equation} 
  Figure 1 illustrates this. At $ \omega = 0 $ the usual London
penetration depth $ \lambda_L$ is found. As the frequency increases $ \lambda_L^* (\omega) $ grows, up to a frequency 
$ \omega_{c 1}, $ where $ k^2 (\omega_{ c 1}) = 0 $ and 
$\lambda_L^* (\omega_{c 1} )$ is infinite, so the field is
uniform inside the medium. For $ \omega_{c1} < \omega < \omega_{T0}, \ k^2 > 0 $ and the field propagates in the medium as in a
dielectric. For $ \omega_{T0} < \omega < \omega_{L 0}, \; k^2 < 0 $ and again there is a frequency dependent $ \lambda_L^* (\omega) ,$ i.e. Meissner effect. Exactly at $ \omega_{L 0}, \ \varepsilon (\omega_{L 0}) = 0, $ so $ \lambda_L^*
(\omega_{LO}) $ coincides with the London penetration depth. For $ \omega_{L0} < \omega < \omega _{c 2}, \ k^2 < 0 $ giving increasing $ \lambda_L^* (\omega) $, and at $ k^2 (\omega_{c 2} ) = 0,$ a uniform field. Then, for $ \omega > \omega_{c 2}, \ k^2 > 0  $ and the wave propagates in the medium again,
as in a dielectric.

In summary, our dispersion equation (9) as a function of frequency gives alternately regions of Meissner-like frequency-dependent penetration depth $ \lambda_L^* (\omega) $ that are superconducting with the magnetic field $ \bf B $ excluded for $ x > \lambda_L^*, $ then changing to regions of field propagation with $ \omega $-dependent real wave number, as in a
dielectric medium. This alternating dielectric and Meissner behavior suggests that our model has a zero-temperature phase transition between normal dielectric and superconductor phases driven by the electromagnetic frequency $ \omega, $ i.e. a type of ''quantum phase transition'' \cite{12}. 

Up to this point damping was ignored . To examine this we write the  dielectric function (4) allowing for a phenomenological
damping coefficient $ \Gamma $ (Ref. \cite{13}):
         \begin{equation} 
 \varepsilon (\omega) = 
  \frac{\varepsilon'_\infty (\omega_{L 0}^2 -
\omega^2) }{(\omega_{T 0}^2 - \omega^2) - 2 i \omega \Gamma}.   
                            \end{equation} 
\begin{figure}[t]
\begin{center}
\includegraphics[width=8cm,height=7.5cm]{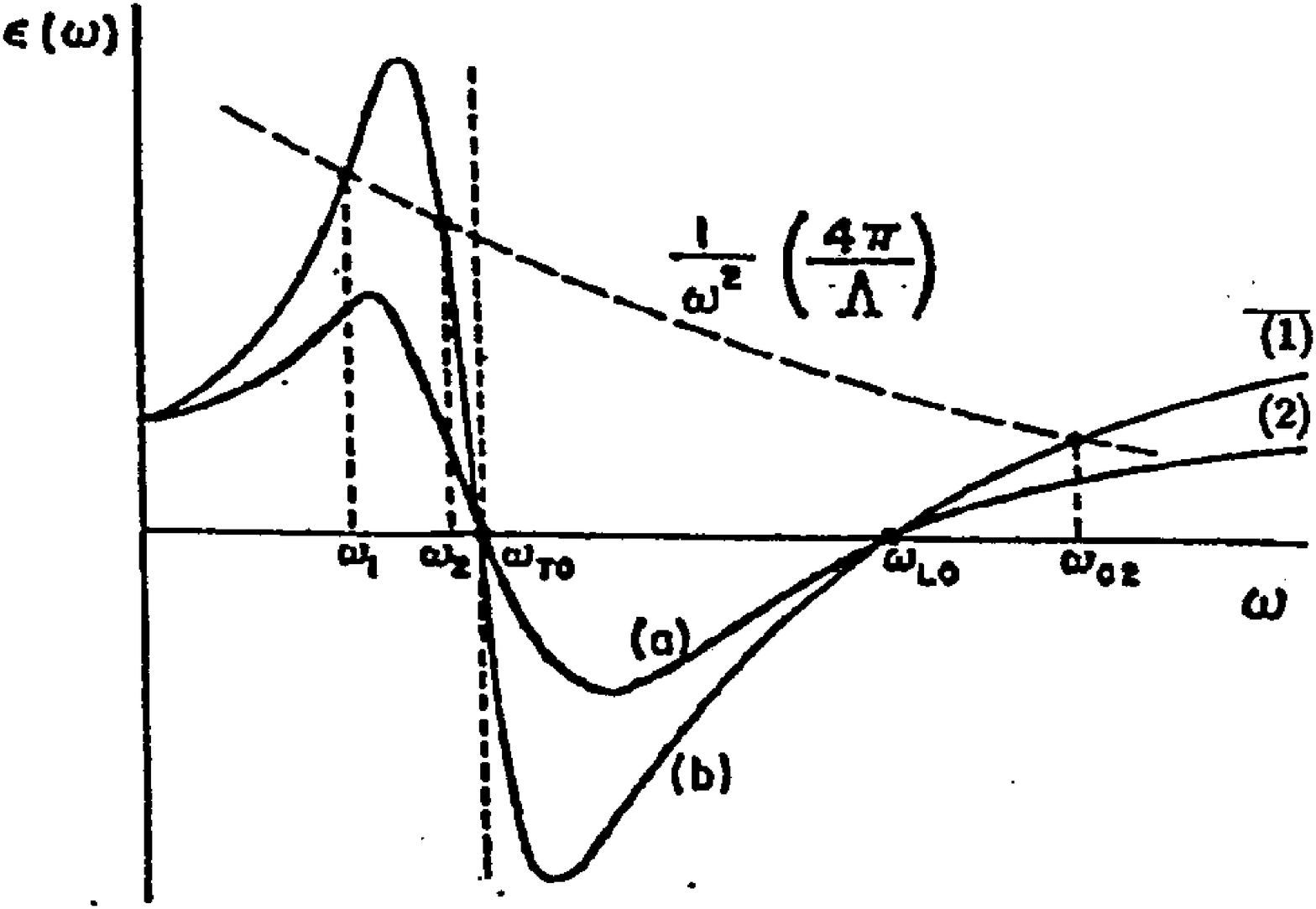}
\caption{ Solutions of the dispersion equation (9) for zero
damping. Frequency intervals $ \omega_{c1} < \omega < \omega_{T 0} $ and $ \omega > \omega_{c 2} $ correspond to real root of the Eq. (9).}
\label{rateI}
\end{center}
\end{figure}

As shown on Fig. 2, including "moderate" damping $\Gamma$ of the soft-mode oscillator changes the results quantitatively but the above qualitative features remain the same.

Our first step is to determine the extent of the important frequency interval $ \omega_{c 1} < \omega < \omega_{T 0}. $ The frequency $ \omega_{c 1} $ is the smallest root of the  Eq. (9) for $ k = 0,$
         \begin{equation}
\varepsilon (\omega) = \frac{4 \pi}{\Lambda} \frac{1}{\omega^2}.
                               \end{equation} 
 The roots of this equation are,
         \begin{equation}
\omega_{1,2} = \omega_{T 0} \sqrt{\frac{\aleph_1}{2} - \frac{2 \Gamma^2 }{\omega_{T 0}^2(1+a^2) } \pm \frac{1}{2} 
\sqrt{\aleph_2 - \frac{4 \Gamma^2 \aleph_1 }{\omega_{T 0}^2 (1+a^2)}  }}. 
                               \end{equation} 
 Here 
     \[
   \aleph_1 = 1 + \frac{1}{1 + a^2}, \ \
\aleph_2 = 1 - \frac{1}{1 + a^2} , \ \
 a^2 = \frac{\omega_{L 0}^2 \varepsilon'_\infty \Lambda }{ 4 \pi }
   \] 
 is a dimensionless parameter which can take on values of the order of unity in nearly ferroelectric materials such as n-SrTiO$ _3 $ (see below).

As we demonstrate below, when we calculate the actual size of the region $ \delta\omega =\omega_{TO} -\omega_{c1} $ for n-SrTiO$_3$ using measured parameters including $\Gamma  , \ \delta\omega$ is, we believe, sufficiently wide to enable measurement of our predicted effects in a film of NFE-SC. First we consider a semi-infinite medium, then the film.

\section{ III. Semiinfinite medium}

We consider a semiinfinite NFE-SC medium occupying the half-space $ z > 0. $ An incident electromagnetic wave propagates in the $ z $ direction with electric and magnetic fields $\bf E (r) $ and $ \bf B (r) $ polarized along $ x $ and $ y $ axes respectively. The Maxwell boundary conditions at the interface plane $ z = 0 $ are \cite{14}
    \bea 
 && E_{i x} (0) + E_{r x} (0) - E_{t x} (0) = 0;
 \nn \\ \nn \\ &&
E_{i x} (0) - E_{r x} (0) - \frac{4 \pi}{c Z} E_{t x} (0) = 0.
                               \eea 
 Here $ E_{ix} (0), E_{rx} (0), E_{tx} (0) $ are the incident, reflected and transmitted electric fields, respectively and $ Z $  is the surface impedance. First consider the case without damping $( \Gamma = 0), $ where we have
         \begin{equation}
Z = \frac{4 \pi}{c} \,\frac{E_{tx} (0) }{B_{ty} (0) }= \mp
\frac{8i\omega}{c^2} (\lambda_L^*)^2
\int_0^\infty \frac{d k}{1 \pm (\lambda_L^* k)^2}.
                               \end{equation} 
 We defined the effective penetration depth $ \lambda_L^* $ by Eq. (10). In the expression (15) the upper signs are used for frequency ranges where the material exhibits Meissner-like behavior $ (k^2 < 0) ,$ while the lower signs apply to the frequency regime of dielectric-like behavior $ (k^2 > 0). $ After integrating over $ k,$ we obtain
 \[
Z = - \frac{4 \pi i \omega}{c^2}\lambda_L^* 
\quad  \mbox{when} \ \omega < \omega_{c 1} \ 
\mbox{or} \ \omega_{T 0} < \omega < \omega_{c 2}  ;\ \ \ 
                       \]
 and
         \begin{equation}
Z =  \frac{4 \pi  \omega}{c^2}\lambda_L^* 
\quad  \mbox{when} \  \omega_{c 1} < \omega < \omega_{T 0} , \  \mbox{or} \ \omega > \omega_{c 2} .
                               \end{equation} 
 Using these results we can solve the equations (14) and calculate the reflection coefficient for the Meissner-like 
$ ( k^2 < 0)$ or dielectric-like $ (k^2 >0). $ frequency
ranges. We obtain the following: 
         \begin{equation}
R = \left |\frac{E_{rx} (0)}{E_{ix} (0)} \right |^2 =
\left (\frac{k_0 \lambda_L^* - 1}{k_0 \lambda_L^* +1} \right )^2;  \qquad k_0 = \frac{\omega}{c}
                               \end{equation} 
  for the anomalous frequency ranges $ (\omega_{c1} < \omega < \omega_{T 0}; \, \, \omega \\> \omega_{c2}) $; and $ R \equiv 1 $ when $ \omega < \omega_{c1} $ or $ \omega_{T 0} < \omega < \omega_{c2}. $ Of course the latter confirms a well-known result of electrodynamics of superconductors, i.e., when we do not consider vortices, the electromagnetic field cannot penetrate into the volume of a superconducting medium. For the ''anomalous'' intervals, e.g., $ \omega_{c1} < \omega < \omega_{T 0} $, the reflection coefficient tends to unity near the limiting points $ \omega_{c1} $ and $ \omega_{T 0} $
because $ \lambda_L^* $ tends to infinity when $ \omega \to \omega_{c1} $ and $ \lambda_L^* $ tends to zero when $ \omega \to \omega_{T 0}. $ However in the interior of this frequency range $ R $ takes values which are considerably smaller than unity. This means that the electromagnetic field can penetrate into a semiinfinite NFE-SC medium within the ''anomalous'' frequency range where the Meissner effect is suppressed. We
did not include the effect of damping which will be discussed below. The damping effect is small in the practical case of n-SrTi$O_3$.

\section{  IV. Thin film of NFE-SC}

We turn to the case of a slab or film of thickness $ L $ so that the material occupies the  region $ 0 \le z \le L. $  As
previously, we take the electromagnetic field normally incident on the interface at $ z = 0, $ propagating in the $z$ direction with the electric and magnetic fields depending only on $z$ and parallel to $x$ and $y$  respectively. To proceed, we expand the electric field in the medium in a Fourier series, as  done in Refs. \cite{7},
         \bea
E(z)& =& \frac{2}{L} \sum \limits_{N=0}^\infty \left(1 -
\frac{1}{2}\delta_{N 0} \right) E_N \cos (k_N z) 
 \nn \\ \nn \\ & +&
\frac{2}{L} \sum \limits_{N=1} \tilde E_N \sin (k_N z).
                               \eea 
 Here
   \bea 
E_N &=& \int_0^L E (z) \cos (k_N z) d z;
               \nn \\ \nn \\ 
\tilde E_N &=& \int_0^L E (z) \sin (k_N z) d z;
\quad   k_N = \frac{\pi N}{L}.\nn
                       \eea
 Using the dispersion equation (9), we get
  \bea 
   E_N &=& \frac{\lambda_L^{*2}}{\mp \lambda_L^{*2} k_N^2 - 1}
\big[E'(0) - (-1)^N E'(L)\big]; \  \
   \nn \\ \nn \\ 
\tilde E_N &=& \frac{\lambda_L^{*2}}{k_N (\mp \lambda_L^{*2} k_N^2 - 1)} \big[E(0) - (-1)^N E(L)\big].
                               \eea
 The upper sign in the denominators of Eq. (19) is used for those frequency ranges where $ k^2 < 0, $  and the lower
sign corresponds to the frequency ranges where $ k^2 > 0. $

Since the new results emerge in the region $\ \delta\omega
=\omega_{C1}-\omega_{TO} $, when $ k^2 >0 $ we consider this case in this Section [lower sign in (19)], and  for the present we take $\Gamma =0$. In this frequency range the material exhibits  dielectriclike behavior. Substituting Eq. (19) into
Eq. (18) we get
   \[
E_x (0) = \frac{i \omega \lambda_L^*}{c} B_y (0) \cot \left
(\frac{L}{\lambda_L^*} \right ) - 
\frac{i \omega \lambda_L^*}{c} B_y (L) \sin^{-1} \left
(\frac{L}{\lambda_L^*}
\right );
                                   \]
         \begin{equation}
E_x (L) = \frac{i \omega \lambda_L^*}{c} B_y (0) \sin^{-1} \left
(\frac{L}{\lambda_L^*} \right ) - 
\frac{i \omega \lambda_L^*}{c} B_y (L) \cot \left (\frac{L}{\lambda_L^*}
\right ).
                               \end{equation} 
 The solution of this system of equations for the magnetic induction is:
  $$
B_y (0) = \frac{i c}{\omega \lambda_L^*} E_x (0) \cot \left
(\frac{L}{\lambda_L^*} \right ) - 
\frac{i c}{ \omega \lambda_L^*} E_x (L) \sin^{-1} \left
(\frac{L}{\lambda_L^*} \right );
             $$
         \begin{equation}
B_y (L) = \frac{i c}{\omega \lambda_L^*} E_x (0) \sin^{-1} \left
(\frac{L}{\lambda_L^*} \right ) - 
\frac{i c}{ \omega \lambda_L^*} E_x (L) \cot \left
(\frac{L}{\lambda_L^*} \right ).
                               \end{equation} 
 The boundary conditions on the interfaces $ z =0 $ and $ z = L $ can be written as  follows:
   \bea && 
E_{ix} (0) + E_{rx} (0) = E_{t x} (0);
         \nn \\ \nn \\ &&
E_{ix} (0) - E_{rx} (0) =  B_y (0);
 \nn \\  \nn \\ &&
E_x (L) = B_y(L).
                               \eea
 Using these boundary conditions and expressions (21) for the magnitude of the magnetic field on the interfaces $ z = 0 $  and $ z = L $ we arrive at the following expressions for the reflection and transmission coefficients:
         \bea &&
R = \frac{1}{1 + \rho^2 (\omega) \sin^{-2} (L/\lambda_L^*)}
                ;    \\ \nn \\ &&
 T = \frac{\rho^2 (\omega) \sin^{-2} (L/\lambda_L^*)}{1 + \rho^2
(\omega) \sin^{-2} (L/\lambda_L^*)},
                               \eea 
 where the the function $ \rho (\omega) $ has the form:
         \begin{equation}
\rho (\omega) = \frac{2 c k^*}{\omega} 
\frac{1}{1 - c^2 k^{* 2}/\omega^2 }.
                               \end{equation} 
 Here $ k^* = 1/\lambda_L^*$ is the solution of the dispersion equation (9) in the ''anomalous'' frequency range $ \omega_{C1} <\omega < \omega_{TO} $. For frequencies very close to 
$ \omega_{T 0} \;$ we have $ k \cong ({\omega}/{c})\sqrt
\varepsilon  $ which corresponds to a polariton-phonon transverse wave.

We now note an interesting result following from Eqs. (23) and
(24), when the effective London penetration depth $\lambda_L^* $, and the Fourier component  $ k=k_N $ are in resonance. That is, for $\lambda_L^* = L/(\pi N)$ or $ k^ = k_N, \ \sin (L/\lambda^*) $ in Eqs. (23) and (24) equals zero and we
get for the reflection and transmission coefficients the values $ R = 0; \ T = 1, $ which means total transparency of the slab at $ k_N $ (See Fig. 3).

Total transparency can also occur at a frequency for which
$ k = k_0 =\omega /c $. At this frequency we have  $ \varepsilon (\omega) - 4 \pi/\omega^2  \Lambda = 1 $ so it is located close to the left boundary frequency $ \omega_{c 1}. $ When  $ k_0 = k $ the function $ \rho(\omega) $ tends to
infinity and again we get $ R = 0 $ and $ T = 1 . $ To explain this we can treat the quantity $ \sqrt{\tilde \varepsilon (\omega)} = \sqrt{\varepsilon (\omega) - 4 \pi/\omega^2 \Lambda} \equiv \tilde n (\omega) $ as an effective reflective index of the slab material. When it equals unity the electromagnetic field from outside passes through the
slab without reflection.

 \begin{figure}[t]
\begin{center}
\includegraphics[width=8cm,height=9cm]{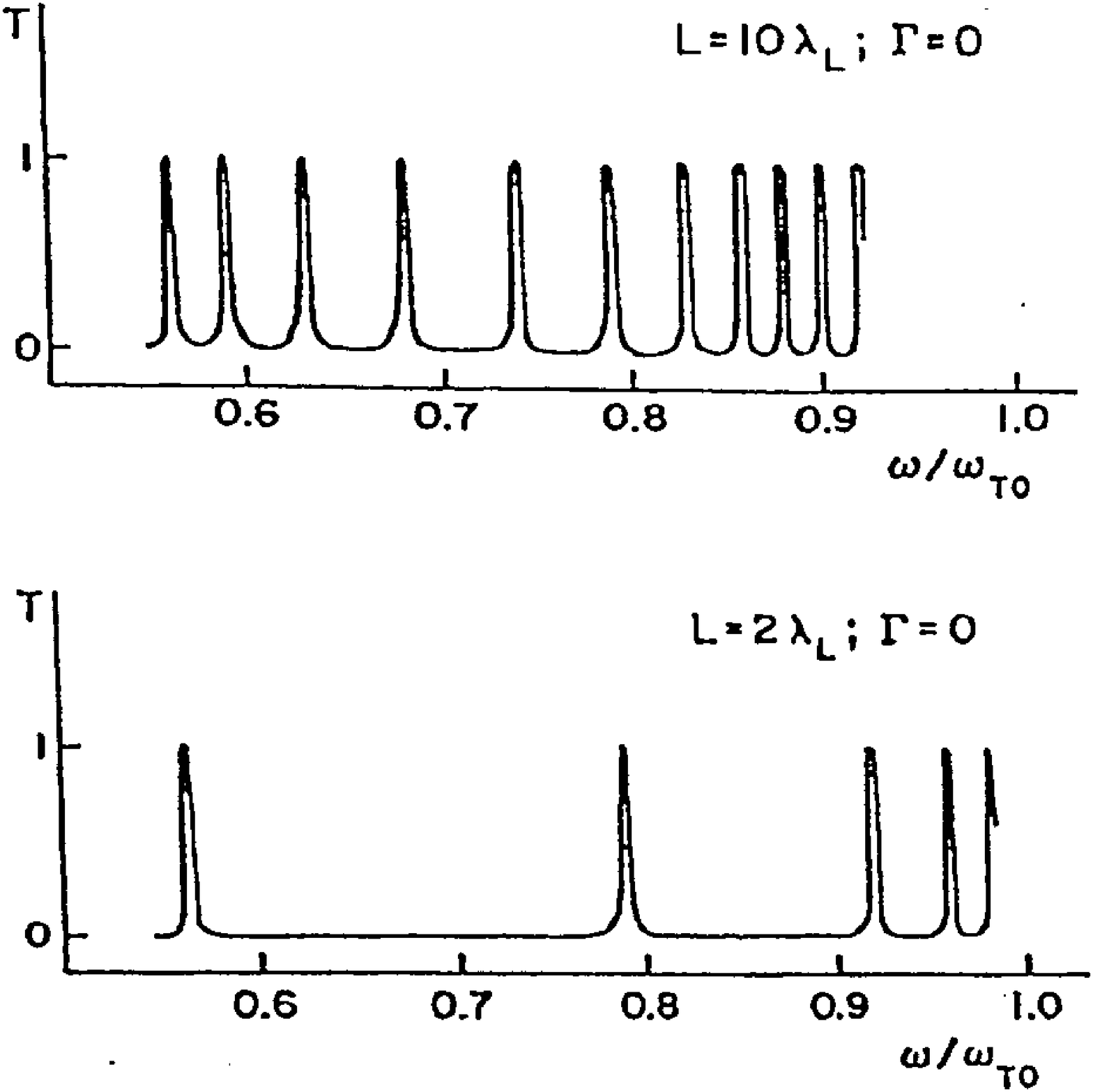}
\caption{ Solutions of the dispersion equation (9) for zero
damping. Frequency intervals $ \omega_{c1} < \omega < \omega_{T 0} $ and $ \omega > \omega_{c 2} $ correspond to real root of the Eq. (9).}
\label{rateI}
\end{center}
\end{figure}

Our results (23) and (24) were obtained, neglecting damping. For a nonzero damping constant $ \Gamma $ we have to replace our expressions (23) and (24) near the resonances by the following:
         \bea
R = \frac{1}{1 + \rho^2 (\omega)+ S^2 (\omega) \sin^{-2}
(L/\lambda{_L^*}')};
                           \\ \nn \\
T = \frac{\rho^2 (\omega) \sin^{-2} (L/\lambda{_L^*}')}{1 + \rho^2 (\omega) + S^2(\omega) \sin^{-2} (L/\lambda{_L^*}')},
                               \eea
 where

         \begin{equation}
S (\omega) = \rho (\omega) \cos \left(\frac{L}{\lambda{_L^*}'} \right ) + \sqrt{1+\rho^2
(\omega)} \sinh \left(\frac{L}{\lambda{_L^*}''} \right) 
                              \end{equation} 
 and $1/\lambda_L^*  = 1/\lambda{_L^*}' + i /\lambda{_L^*}''. $

We can use the following approximation for $ \lambda{_L^*}' $ and $ \lambda{_L^*}'' :$
         \bea  &&
\frac{1}{\lambda{_L^*}'} \approx \sqrt{\frac{\omega^2}{c^2} \varepsilon ' (\omega) - \frac{1}{\lambda_L^2}};
             \\ \nn \\ &&
\frac{1}{\lambda{_L^*}''} \approx \left (\frac{2 \omega \Gamma}{\omega_{T 0}^2 - \omega^2} \right )^{1/2} \frac{1}{\lambda_L}.
                               \eea 
 It was shown before that damping in the lattice system reduces
the anomalous frequency range $ \delta \omega .$ The frequency interval $ \omega_{c 1} < \omega < \omega_{T 0} $ has to be replaced by the narrower interval $ \omega_{ 1} < \omega < \omega_{2} .$ The imaginary part of $ 1/\lambda{_L^*} $ depends on frequency and increases when the latter increases. As we will illustrate below, moderate damping, as in  n-SrTiO$ _3 $ will only slightly modify the sharp resonances which occur at $k =k_N$ and $k =k_0. $

 To the best of our knowledge the general behavior we predict for an NFE-SC system has not been observed. In order to examine the practicality of making such observations we have calculated the effects quantitatively for n-doped SrTiO$_3$. This material was much studied, as a nearly-ferroelectric, and as a superconductor, prior to the era of high T$_c$, so that the needed physical parameters are available.

\section{ V. Application to n-SrTiO$_3$, a Prototype NFE-SC}

In order to evaluate the magnitudes and locations of the predicted effects, we require lattice, dielectric, and electronic data. So far as we can determine such measured data is only available for n-SrTiO$_3$, and we will use it below. Nearly ferroelectric SrTiO$_3$ has been studied for some time. The temperature dependent static dielectric coefficient 
 $\varepsilon (0)$ has been measured in a wide temperature range, extending down to about 1K [4]. The frequency dependent
 $\varepsilon_{\infty}$ is known at various temperatures [4]. Lattice dynamic studies were carried out based on the shell model [13], and neutron scattering [13], and infra-red and Raman (including electric field dependence) optical properties have been measured [9]. As a result the lattice dynamical parameters are known, and the generalized LST relation (5) has been verified. From the line-widths in infra-red and Raman properties [9], the damping parameter $\Gamma $ for the soft mode lattice frequency $\omega_{TO}$ is known. Hence all the physical parameters of the dielectric host are actually known. The temperature independent quantities are
    \[
   \varepsilon_{\infty} =5.5, \quad
   \omega_{LO} =5.2 \times 10^{12} s^{-1},
            \quad                      
  \frac{ \Pi_{i}'\omega_{Li}^2}{\omega_{Ti}^2 } =4.1 . 
                               \]            
 Using the LST relation we have at $\sim$1K
   \[
  \omega_{TO} =1.6 \times 10^{11} s^{-1}
                                     \]
and at 1K
   \[
  \varepsilon (0)\sim 2.25\times10^4  .
                                  \] 
The damping constant for the soft TO mode as in eqn(13) is measured from optical line width \cite{9}:
   \[
 \Gamma^2/\omega_{T 0}^2 \sim  0.2 
                                \]
 at the temperature of interest. (Actually the measurement was at a higher temperature than the region of T$_c$ so the relevant $\Gamma$ for our work will be smaller, but we use the above value.)

For the needed electronic parameters  we use the  effective mass
$m^* \sim 10m_e $ where $m_e$ is the free electron mass. A range of $n$ doping from $ 10^{17}$ to $10^{21}$ cm$^{-3}$ was used for SC n-SrTiO$_3 $. We choose $n_s =9\times 10^{17}$ cm$^{-3}$ for our calculation. We recall that n-doped SrTiO$_3$ exhibits concentration dependent T$_c$, with T$_c$ = 0.3K in the optimum doped system \cite{3}. In this temperature range $\varepsilon (0)$ has the high value given above, as  was noted also by Cohen \cite{3}.

We now determine the anomalous frequency interval just below
$\omega_{TO}$. We first need the parameter $a^2 =\omega_{LO}^2
\varepsilon_{\infty}'\Lambda /4\pi$ [see Eq. (13)]. For SrTiO$_3$ with $n_s =9 \times10^{17}$ cm$^{-3}$, and $m^* \approx 10m_e $, we get $a^2\approx 2.1$. Taking, first, $\Gamma =0$, and solving Eq. (12) we find $\omega_{c1}=0.9 \times 10^{11} c^{-1}$. This gives $\delta\omega = \omega_{TO} - \omega_{c1} =0.43 \omega_{TO}.$ Using the measured $\Gamma$ for SrTiO$_3$, we obtain that $\omega_2$ shifts negligibly while $\omega_{c1}$ shifts slightly to $\omega_{c1}' =0.96 \times 10^{11}c^{-1} $, and the anomalous frequency region becomes slightly smaller
   \[
  \delta \omega' = \omega_{TO} -\omega_{c1}'\sim 0.4\omega_{TO}.
                                                  \]
We expect that this interval $\delta\omega'$ is still sufficiently wide to examine optical response experimentally by reflection/transmission studies in n-SrTiO$_3$. The small effect of damping makes it likely also that $\Gamma =0$
is a good approximation for the calculations that follow.

We now turn to calculating the reflection/transmission coefficients for a $n$-SrTiO$_3$ film of NFE-SC of thickness $L$, first for $\Gamma =0$, using the formulas (23)-(25). In this case  the results are shown on Fig. 3. We predict a ''comb-like'' structure of narrow perfect transmission $T_\rho=1 $ spikes alternating with regions where $T_\rho=0.$ The spikes occur at the resonance condition where $\lambda_L^* = L/\pi N $. Next we estimate the effect including damping. For weak damping $\Gamma^2/\omega_{TO}^2 =0.2$ the
value of the factor $ \big [2 \omega \Gamma \big /(\omega_{T 0}^2 - \omega^2) \big]^{1/2} $ changes from 0.9 to 1.5 when $ \omega $ runs over the interval between $ \omega_1 $ and $ \omega_2 .$ Thus, when the thickness of our film $ L $ is small or of the same order as the $\omega =0$ London penetration depth $ \lambda_L $ (here $ \lambda_L \sim 1.8 \times
10^{-3} $ cm) moderately weak damping as in SrTiO$_3$ cannot significantly change the height of the peak in the transmission coefficient. Peaks arise due  to the propagation of the transverse mode corresponding to the solution of the dispersion equation (9). We believe that these peaks can be observed
in any NFE-SC material in the  appropriate frequency ranges, and at temperatures below $ T_c. $ Such an observation would give  support for theory proposed here.

It is not possible to make realistic estimates of the interval
$\delta\omega$ (or $\delta\omega '$ with damping) for Na$_x$WO$_3$, or for cuprates pending the availability of the necessary material constants as we had in $n$-SrTiO$_3$. The theory presented will apply to any such NFE-SC material, so the experimental search for the resonance comb-like structure in T($\omega$) just below $\omega =\omega_{TO}$ will be highly
valuable.

\section{VI. Summary}

We solved the coupled "soft-mode" lattice dynamics (long wavelength) equations of motion, together with Maxwell equations for the associated electromagnetic field and the London equations for the supercurrent, in a NFE-SC. The analysis can apply to several material systems of current interest: sodium tungsten bronzes Na$_x$WO$_3$, doped 
$n$-SrTiO$_3$, and possibly high temperature cuprates.
We note that several authors have discussed microscopic theories for the superconductivity in $n$-SrTiO$_3$ (Ref. \cite{15}) in the strong-coupling framework. However, our macroscopic approach has not previously been reported, to
our knowledge. We assume all the free carriers are condensed in Cooper pairs, as in some earlier work on the SrTiO$_3$ (See Regs. \cite{3,4}). The resulting coupled modes can be considered as phonon-polaritons dressed by the supercurrent or electromagnetic waves in a London superconductor
dressed by the transverse optic waves (TO phonons).

A result, which we illustrated, is the alternation of the system
response between Meissner-like superconductor and a dielectric-like medium, as the incident electromagnetic frequency is continuously varied. Particularly of interest is the  "comb-like" series of transmission resonance peaks we predict, which should be measurable. Such measurements would test our theory. One NFE-SC material of choice would be $n$-SrTiO$_3$ for which we made quantitative estimations.

\section{ Acknowledgments}

We thank G.M. Zimbovsky for his help with the manuscript. An
inservice grant from the PSC-CUNY is also acknowledged for partial support.

\section{Appendix: Partition of Energy Between Radiation, 
  Lattice Polarization, and London Field }

It is useful to obtain the frequency-dependent partition of energy between the three propagating fields. This can be seen as extending Huang's early treatment of the radiation-lattice coupling (phonon-polariton \cite{8})  to include the London supercurrent, or conversely extending London treatment
\cite{16} of the energy momentum theorem, to include the dielectric polarization. 

Within our treatment all fields are transverse and we consider the infinite medium. We will obtain the continuity equation for energy density $ U $ and Poynting energy flux vector $ \bf s . $ From Maxwell's equations (1) we have:
  $$
{\bf \nabla \cdot s} = 
- \frac{1}{4 \pi} {\bf H} \cdot \frac{\partial \bf B}{d t}
- \frac{1}{4 \pi} {\bf E} \cdot \frac{\partial \bf D}{d t},
                  \eqno(A.1)$$
  where 
$$ \bf s =\frac{c}{4 \pi} [E \times H].
                  \eqno(A.2)$$
                       
In order to simplify the right hand side of (A1) we use the lattice equation of motion (3) to obtain:
  $$ 
\frac{1}{c}{\bf E} \cdot \frac{\partial \bf D}{d t} = \frac{1}{2 c}
\varepsilon_\infty \frac{\partial}{\partial t} ({\bf E}^2 ) +
\frac{2\pi}{c} \frac{\partial}{\partial t} 
\left (\frac{\partial}{\partial t} {\bf w}^2 + \omega_{T0}^2 {\bf w}^2
\right ).
                  \eqno(A.3)$$
 Then, using the first of equations (2), and equations (6) and (7) we have
  $$ 
{\bf [\nabla \times M]} = \frac{1}{c} {\bf J} = - \frac{1}{c^2
\Lambda} \bf A
                  \eqno(A.4)$$
 and
  $$
{\bf[\nabla \times [\nabla \times M]]} = - \nabla^2 {\bf M} = -
\frac{1}{c^2 \Lambda} \bf B.
                  \eqno(A.5)$$

Now we will assume plane monochromatic electromagnetic wave propagation $ ( \sim \exp (i {\bf k \cdot r} - i \omega t)).$ So
        $$
{\bf \nabla}^2 \to - k^2; \qquad \qquad \frac{\partial}{\partial t} \to -
i \omega.
                $$
 Then
  $$
k^2 {\bf M} = - \frac{1}{c^2 \Lambda} {\bf B},
                  \eqno(A.6)$$
 so
  $$
{\bf B = H} - \frac{1}{\lambda_L^2 k^2} \bf B,
                  \eqno(A.7)$$
or with $ \bf B = {\mu} H $ we get:
  $$
\frac{1}{\mu} = 1 + \frac{1}{\lambda_L^2 k^2}
                  \eqno(A.8)$$
 and
  $$
- \frac{1}{c} {\bf H} \cdot \frac{\partial \bf B}{\partial t} = -  \frac{1}{c} \frac{1}{\mu} {\bf B} \frac{\partial \bf B}{\partial t} = - \frac{1}{2c} \left (1 + \frac{1}{\lambda_L^2 k^2} \right ) \frac{\partial}{\partial t} {\bf B}^2.
                  \eqno(A.9)$$
 Since all waves are transverse take $ {\bf E} = (E_x, 0,0); \ {\bf B} = (0, B_y,0); \ {\bf k} = (0,0,k) $ giving $B_y =
({c k}/{\omega}) E_x.$ We eliminate $ \bf H $ from (A.1), and combining all terms, we get
  $$
\frac{\partial U}{d t} = \frac{1}{4\pi} {\bf H} \cdot \frac{\partial \bf B}{\partial t} + \frac{1}{4 \pi} {\bf E} \frac{\partial \bf D}{\partial t}
= \frac{1}{8 \pi} \left (\varepsilon_\infty \frac{\partial}{\partial t}
{\bf E}^2 + \frac{\partial {\bf B}^2}{\partial t} \right ) 
                  $$
  $$ -
\frac{1}{2} \frac{\partial}{\partial t} {\bf {\dot w}}^2 +
\frac{\omega_{T0}^2}{2} \frac{\partial}{\partial t} {\bf w}^2 +
\frac{1}{8 \pi} \frac{c^2}{\omega^2 \lambda_L^2} \frac{\partial}{\partial t} {\bf E}^2.
                  \eqno(A.10)$$
 It is simple to identify the various terms in the energy density $ U $ as belonging to radiation field, lattice polarization, and London supercurrent. Thus
  $$
U_{RAD} = \frac{1}{8 \pi} \varepsilon_\infty {\bf E}^2 + \frac{1}{8 \pi} {\bf B}^2,
                  \eqno(A.11)$$
      $$
U_{LATTICE} =-\frac{1}{2} {\bf{\dot w}}^2 + 
\frac{1}{2}\omega_{TO}^2 {\bf w}^2 ,
          \eqno(A.12)$$
  $$
U_{SC} =\frac{1}{8\pi}\frac{1}{\lambda_{L}^2 k_{0}^2}{\bf E}^2, 
  \eqno(A.13)$$
 where $k_{0}= {\omega}/{c}$. All these contributions to
the energy density can be expressed in terms of ${\bf E}^2$ in our plane-wave case, giving
    $$
U_{RAD} = \frac{1}{4\pi}\bigg(\varepsilon_{\infty}+ \frac{k^2}{k_{0}^2}\bigg){\bf E}^2 ,
          \eqno(A.14)$$
  $$
U_{LATTICE} = \frac{(\varepsilon_{0} -\varepsilon_{\infty})\omega_{TO}^2 (\omega^2 +\omega_{TO}^2)}{8 \pi (\omega_{TO}^2 -\omega^2)^2}{\bf E}^2 
           \eqno(A.15)$$
 and, as before:
  $$
U_{SC} = \frac{1}{8\pi}\frac{1}{\lambda_{L}^2k_{0}^2} {\bf E}^2. 
     \eqno(A.16)$$

Before proceeding we note that the expression for the lattice term in energy density is unchanged from that found by Huang \cite{8} for the phonon-polariton. Formally that is also true for the radiative term, however hidden in $U_{RAD}$ is the effect of the London term via the changed dispersion equation (9), which includes $\Lambda$. Note that substituting the dispersion equation (9) into $U_{RAD}$ we get
   $$
U_{RAD} = \frac{1}{8\pi} \left(\varepsilon_{\infty} +\varepsilon(\omega) -
\frac{1}{\lambda_{L}^2k_{0}^2} \right){\bf E}^2
       \eqno(A.17)$$
 or
  $$
U_{RAD} = \frac{1}{8\pi} \left( 2\varepsilon_{\infty} +
\frac{\varepsilon_{\infty}(\omega_{LO}^2 -\omega_{TO}^2)}{\omega_{TO}^2 -
\omega^2} -\frac{1}{\lambda_{L}^2 k_{0}^2}\right){\bf E}^2 ,
 \eqno(A.18)$$
  and
  $$
U_{SC} = \frac{1}{8\pi}\frac{1}{\lambda_{L}^2 k_{0}^2} {\bf E}^2 .
                            \eqno(A.19)$$
 We may see that the energy-density contribution $U_{SC}$ will exactly cancel the London contribution to the radiation energy density $U_{RAD}$. Thus, if we calculate the lattice (polarization) fraction of the energy we obtain exactly the result of Huang \cite{8} despite the coupling of the London supercurrent. We have
  \bea
\rho_{LATTICE}& =&\frac{U_{LATTICE}}{U_{TOTAL}} =
\frac{U_{LATTICE}}{U_{LATTICE} + U_{RAD} + U_{SC}}
   \nn \\ \nn \\& =&
\frac{1}{2}\frac{(\omega_{TO}^2 +\omega^2)(\omega_{LO}^2 -
\omega_{TO}^2)}{(\omega_{TO}^2 - \omega^2)^2 +\omega_{TO}^2 (\omega_{LO}^2
-\omega^2)} . \ \ (A.20) \nn
 \eea

This result shows that the mode described by the dispersion equation (9) is a mixed photon-phonon mode within all frequency range $\omega_{c1} < \omega <\omega_{TO}$. We emphasize this result is only meaningful in the intervals where $k^2 > 0$. The lattice contribution to the energy enhances
when the frequency increases and gets its maximum when $\omega$ goes to $ \omega_{TO}$. At the frequencies close to $ \omega_{TO}$ $\rho\approx 1$, so all the energy density is concentrated in the ion system. Taking into account that the frequency $\omega_{TO}$ is a boundary frequency for the
"anomalous" frequency range, we can expect that the ion oscillations at the frequency $\omega_{TO}$ being excited by the electromagnetic field which can penetrate to a NFE-SC material within the "anomalous" frequency interval, will exist at frequencies larger than $\omega_{TO}$ where the
material behaves as a superconductor.

\end{document}